\documentclass[preprint,showpacs,preprintnumbers,amsmath,amssymb]{revtex4}

\usepackage{epsfig}
\usepackage{graphicx}
\usepackage{dcolumn}
\usepackage{bm}

\newcommand{\ord}{{\cal O}}
\def\beq{\begin{equation}}
\def\eeq#1{\label{#1}\end{equation}}
\def\eeqn{\end{equation}}
\newcommand\iden{\leavevmode\hbox{\small1\normalsize\kern-.33em1}}

\let\jnfont=\rm
\def\RMP#1,{{\jnfont Rev.\ Mod.\ Phys }{\bf #1},}
\def\NPB#1,{{\jnfont Nucl.\ Phys.\ B }{\bf #1},}
\def\PLB#1,{{\jnfont Phys.\ Lett.\ B }{\bf #1},}
\def\EPJC#1,{{\jnfont Eur.\ Phys.\ Jour.\ C }{\bf #1},}
\def\PRD#1,{{\jnfont Phys.\ Rev.\ D }{\bf #1},}
\def\PRL#1,{{\jnfont Phys.\ Rev.\ Lett.\ }{\bf #1},}
\def\MPLA#1,{{\jnfont Mod.\ Phys.\ Lett.\ A }{\bf #1},}
\def\JPG#1,{{\jnfont J.\ Phys.\ G }{\bf #1},}
\def\CTP#1,{{\jnfont Commun.\ Theor.\ Phys.\ }{\bf #1},}
\def\JHEP#1,{{\jnfont JHEP \ }{\bf #1},}
\def\NPPS#1,{{\jnfont Nucl.\ Phys.\ Proc.\ Suppl.\ }{\bf #1},}
\def\CPC#1,{{\jnfont Computl.\ Phys.\ Commun.\ }{\bf #1},}

\newcommand{\be}{\begin{equation}}
\newcommand{\ee}{\end{equation}}
\newcommand{\bea}{\begin{eqnarray}}
\newcommand{\eea}{\end{eqnarray}}
\begin{document}

\preprint{\parbox{1.2in}{\noindent ~}}

\title{\ \\[10mm]  Higgs and Z-boson FCNC decays correlated with B-meson decays \\
                   in littlest Higgs model with T-parity}

\author{\ \\[2mm] Xiao-Fang Han, Lei Wang, Jin Min Yang \\ ~}

\affiliation{Institute of Theoretical Physics, Academia Sinica,
             Beijing 100080, China \vspace*{1.5cm}}

\begin{abstract}
In the littlest Higgs model with T-parity (LHT) new flavor-changing
interactions between mirror fermions and the Standard Model (SM)
fermions can induce various FCNC decays for B-mesons, Z-boson and
Higgs boson. Since all these decays induced in the LHT model are
correlated, we in this work perform a collective study for these
decays, namely the Z-boson decay $Z \to b \bar s $, the Higgs boson
decay $h \to b \bar s$, and the B-meson decays $B \to X_s \gamma$,
$B_s \to \mu^+\mu^-$ and $B \to X_s \mu^+ \mu^-$. We find that under
the current experimental constraints from the B-decays, the
branching ratios of both $Z \to b \bar{s}$ and $h \to b \bar{s}$ can
still deviate from the SM predictions significantly. In the
parameter space allowed by the B-decays, the branching ratio of $Z
\to b \bar{s}$ can be enhanced to $10^{-7}$ (about one order above
the SM prediction) while $h\to b \bar{s}$ can be much suppressed
relative to the SM prediction (about one order below the SM
prediction).

\vspace*{1cm}
\end{abstract}


\maketitle

\section{Introduction}
The fancy idea of little Higgs \cite{ref1} tries to provide an
elegant solution to the hierarchy problem by regarding the Higgs
boson as a pseudo-Goldstone boson whose mass is protected by an
approximate global symmetry and free from one-loop quadratic
sensitivity to the cutoff scale. The littlest Higgs model
\cite{ref2} is a cute economical implementation of the little Higgs
idea, but is found to be subject to strong constraints from
electroweak precision tests \cite{ref3}, which would require raising
the mass scale of the new particles to far above TeV scale and thus
reintroduce the fine-tuning in the Higgs potential \cite{ref4}. To
tackle this problem, a discrete symmetry called T-parity is proposed
\cite{ref5}, which forbids the tree-level contributions from the
heavy gauge bosons to the observables involving only SM particles as
external states. However, in such littlest Higgs model with T-parity
(LHT) \cite{ref5}, there arise new flavor-changing interactions
between mirror fermions and the SM fermions (just like the
flavor-changing interactions between sfermions and fermions in
supersymmetric models). Such new flavor-changing interactions can
have crucial phenomenology, especially they can induce various
flavor-changing neutral-current (FCNC) processes, which should be
examined.

Among various FCNC processes induced by the new flavor-violating
interactions in the LHT model, the loop-induced B-decays, such as $B
\to X_s \gamma$, $B_s \to \mu^+\mu^-$ and $B \to X_s \mu^+\mu^-$,
should be first checked due to the available experimental data on
these decays. Recently, these B-decays have been intensively
examined in the LHT model, which were found to be sensitive to the
new flavor-violating interactions \cite{blht1,blht2,blht3}. Note
that in addition to these B-decays, the loop-induced FCNC decays of
Higgs and Z-boson, such as $Z \to b \bar{s}$ and $h \to b \bar{s}$
which are strongly correlated with the FCNC B-decays, should also be
examined since they are sensitive to the flavor structure of new
physics.  In the future there may be at least two avenues in which
Z-bosons will be produced in much larger quantities than at LEP. At
the CERN Large Hadron Collider (LHC) with an integrated luminosity
of 100 fb$^{-1}$, one expects $5.5 \times 10^9$ Z-bosons to be
produced \cite{LHC}. In particular, the GigaZ option at the proposed
International Linear Collider (ILC) with an integrated luminosity of
30 fb$^{-1}$, it is possible to produce more than $10^9$ Z-bosons
\cite{ILC}. For the study of the Higgs boson, one may expect the ILC
to scrutinize the Higgs boson property after the discovery at the
LHC.

These rare decays $Z \to b  \bar{s}$ and $h \to b \bar{s}$  have
been studied in the SM \cite{maxzsbsm} and in various new physics
models \cite{zbsnp,lbsnp}. In this work we will study these decays
in the LHT model. Since such decays are strongly correlated with the
induced FCNC B-decays ( $B \to X_s \gamma$, $B_s \to \mu^+\mu^-$ and
$B \to X_s \mu^+\mu^-$), we will collectively consider all these
decays. We will first check the analytic results of these B-decays
given in \cite{blht1,blht2,blht3} and then perform their numerical
calculations together with $Z \to b  \bar{s}$ and $h \to b \bar{s}$.
We will show the constraints on the parameter space from current
B-decay experiments and display the results for  $Z \to b  \bar{s}$
and $h \to b \bar{s}$ with/without the B-decay constraints.

This work is organized as follows. In Sec. II we recapitulate the
LHT model and address the new flavor violating interactions which
will contribute to the FCNC decays considered in this work.
In Sec. III , IV and V we examine the B-decays ($B \to X_s \gamma$,
$B_s \to \mu^+\mu^-$ and $B \to X_s \mu^+\mu^-$), Z-boson decay $Z \to b \bar{s}$
and Higgs boson decay  $h \to b \bar{s}$, respectively.
Finally, we give our conclusion in Sec. VI.

\section{The littlest Higgs model with T-parity}

The LHT model \cite{ref5} is based on a non-linear sigma model
describing the spontaneous breaking of a global $SU(5)$ down to a
global $SO(5)$ by a 5$\times$5  symmetric tensor at the scale
$f\sim\ord(TeV)$. From the $SU(5)/SO(5)$ breaking, there arise 14
Goldstone bosons which are described by the "pion" matrix $\Pi$,
given explicitly by
\small
\be \label{Pi}
 \addtolength{\arraycolsep}{3pt}\renewcommand{\arraystretch}{1.3}
 \Pi=\left(\begin{array}{ccccc}
-\frac{\omega^0}{2}-\frac{\eta}{\sqrt{20}} &
-\frac{\omega^+}{\sqrt{2}} &
  -i\frac{\pi^+}{\sqrt{2}} & -i\phi^{++} & -i\frac{\phi^+}{\sqrt{2}}\\
-\frac{\omega^-}{\sqrt{2}} &
\frac{\omega^0}{2}-\frac{\eta}{\sqrt{20}} &
\frac{v+h+i\pi^0}{2} & -i\frac{\phi^+}{\sqrt{2}} & \frac{-i\phi^0+\phi^P}{\sqrt{2}}\\
i\frac{\pi^-}{\sqrt{2}} & \frac{v+h-i\pi^0}{2} &\sqrt{4/5}\eta &
-i\frac{\pi^+}{\sqrt{2}} & \frac{v+h+i\pi^0}{2}\\
i\phi^{--} & i\frac{\phi^-}{\sqrt{2}} & i\frac{\pi^-}{\sqrt{2}} &
-\frac{\omega^0}{2}-\frac{\eta}{\sqrt{20}} & -\frac{\omega^-}{\sqrt{2}}\\
i\frac{\phi^-}{\sqrt{2}} &  \frac{i\phi^0+\phi^P}{\sqrt{2}} &
\frac{v+h-i\pi^0}{2} & -\frac{\omega^+}{\sqrt{2}} &
\frac{\omega^0}{2}-\frac{\eta}{\sqrt{20}}
\end{array}\right).
\ee 
\normalsize 
Under T-parity the SM Higgs doublet $ H= \left(-i
\pi^+/ \sqrt{2}, (v+h+i\pi^0)/2 \right)^T$ is T-even, while the
other fields are T-odd. A subgroup $[SU(2)\times
U(1)]_{1}\times[SU(2)\times U(1)]_{2}$ of the $SU(5)$ is gauged, and
at the scale $f$ it is broken into the SM electroweak symmetry
$SU(2)_L\times U(1)_Y$. The Goldstone bosons $\omega^{0}$,
$\omega^{\pm}$ and $\eta$ are respectively eaten by the new T-odd
gauge bosons $Z_{H}$, $W_{H}$ and $A_{H}$, which obtain masses at
$\ord$$(v^2/f^2)$ \be M_{W_H}=
M_{Z_H}=fg\left(1-\frac{v^2}{8f^2}\right), ~~ M_{A_H}=\frac{f
g'}{\sqrt{5}}\left(1-\frac{5v^2}{8f^2}\right), \ee with $g$ and
$g^\prime$ being the SM $SU(2)$ and $U(1)$ gauge couplings,
respectively.

The  Goldstone bosons $\pi^{0}$ and $\pi^{\pm}$ are eaten by the SM
T-even $Z$-boson and $W$-boson, which obtain masses at $\ord(v^2/f^2)$,
\be 
M_{W_L}=\frac{gv}{2}\left(1-\frac{v^2}{12f^2}\right),\quad
M_{Z_L}=\frac{gv}{2\cos\theta_W}\left(1-\frac{v^2}{12f^2}\right).
\ee
The photon $A_L$ is also T-even and massless. Due to the mass of
SM bosons corrected at $\ord(v^2/f^2)$, the relation between
$G_F$ and $v$ is modified from its SM form and is given by
$\frac{1}{v^2}=\sqrt{2}G_{F}(1-\frac{v^2}{6f^2})$.

The top quark has a T-even partner T quark and a T-odd $T_-$ quark.
To leading  order, their masses are given by 
\be 
 M_{T}=\frac{m_tf}{v}(r+\frac{1}{r}),\quad
 M_{T_-}=M_{T}\frac{1}{\sqrt{1+r^2}},
\ee 
where $r=\lambda_1/\lambda_2$ with $\lambda_1$ and $\lambda_2$
being the coupling constants in the Lagrangian of the top quark
sector \cite{ref5}. Furthermore, for each SM quark (lepton), a copy
of mirror quark (lepton) with T-odd quantum number is added in order
to preserve the T-parity. We denote them by $u_H^i, d_H^i,
\nu_H^i, l_H^i$, where $i=1, 2, 3$ are the generation index. In
$\ord(v^2/f^2)$ their masses satisfy 
\be
m_{d_H^i}=\sqrt{2}\kappa_{q^i}f,\qquad
m_{u_H^i}=m_{d_H^i}(1-\frac{v^2}{8f^2}). \label{eq5} 
\ee   
Here $\kappa_{q^i}$ are the diagonalized Yukawa couplings of the mirror quarks.

Note that new flavor interactions arise between the mirror fermions
and the SM fermions, mediated by the T-odd gauge bosons or T-odd
Goldstone bosons. In general, besides the charged-current
flavor-changing interactions, the FCNC interactions between the
mirror fermions and the SM fermions can also arise from the mismatch
of rotation matrices. For example, there exist FCNC interactions
between the mirror down-type quarks and the SM down-type quarks,
where the mismatched mixing matrix is denoted by $V_{H_{d}}$. We
follow \cite{blht1,blht3} to parameterize $V_{H_{d}}$ with three
angles $\theta_{12}^d,\theta_{23}^d,\theta_{13}^d$ and three phases
$\delta_{12}^d,\delta_{23}^d,\delta_{13}^d$ 
\small 
\be
\left(\begin{array}{ccc} c_{12}^d c_{13}^d & s_{12}^d c_{13}^d
e^{-i\delta^d_{12}}& s_{13}^d e^{-i\delta^d_{13}}\\
-s_{12}^d c_{23}^d e^{i\delta^d_{12}}-c_{12}^d s_{23}^ds_{13}^d
e^{i(\delta^d_{13}-\delta^d_{23})} & c_{12}^d c_{23}^d-s_{12}^d
s_{23}^ds_{13}^d e^{i(\delta^d_{13}-\delta^d_{12}-\delta^d_{23})} &
s_{23}^dc_{13}^d e^{-i\delta^d_{23}}\\
s_{12}^d s_{23}^d e^{i(\delta^d_{12}+\delta^d_{23})}-c_{12}^d
c_{23}^ds_{13}^d e^{i\delta^d_{13}} & -c_{12}^d s_{23}^d
e^{i\delta^d_{23}}-s_{12}^d c_{23}^d s_{13}^d
e^{i(\delta^d_{13}-\delta^d_{12})} & c_{23}^d c_{13}^d
\end{array}\right).
\ee
\normalsize

\section{FCNC B-decays}
The decays $B \to X_s \gamma$, $B_s \to \mu^+\mu^-$  and  $B \to X_s
\mu^+\mu^-$ can be induced at loop level by the new flavor-changing
interactions in the LHT model and have been recently studied in
\cite{blht3}. We check the results of \cite{blht3} and make the
following brief descriptions about these decays without giving the
detailed expressions (all functions in our following discussions can
be found in \cite{blht3}).
\begin{itemize}
\item[(1)] For $B \to X_s \gamma$ the LHT contributions enter through
the modifications of the quantities
\be
T_{D'}^\text{SM} \equiv \lambda_t^{(s)} D_0'(x_t)
  =-2\lambda_t^{(s)}C^{0^{SM}}_{7\gamma}(M_W),~~
T_{E'}^\text{SM} \equiv \lambda_t^{(s)}
E_0'(x_t)=-2\lambda_t^{(s)}C^{0^{SM}}_{8G}(M_W), 
\ee 
where the CKM
factor $\lambda_t^{(s)}=V_{ts}V_{tb}^*$, and $C^{0^{SM}}_{7\gamma}$
and $C^{0^{SM}}_{8G}$ are leading-order Wilson coefficients. With
the LHT effects $T_{D'}^\text{SM}$ and $T_{E'}^\text{SM}$ are
replaced by $T_{D'}$ and $T_{E'}$ 
\be 
  T_{D'}=T_{D'}^{\rm even}+T_{D'}^{\rm odd}, 
~~T_{E'}=T_{E'}^{\rm even}+T_{E'}^{\rm odd},
\ee 
where the superscripts 'even' and 'odd' denote the contributions
from T-even and T-odd particles, respectively. Note that for the LHT
contributions we only consider the leading-order effects while for
the SM prediction we consider the next-to-leading-order QCD
corrections. Actually, the SM prediction for $B \to X_s \gamma$ has
been calculated to the NNLO \cite{bsr2loop}.

\item[(2)]  The branching ratio of $B_s \to  \mu^+\mu^-$ in the SM depends
on a function $Y_{SM}$ and the LHT effects enter through the
modification of $Y_{SM}$ \cite{blht3}. With the LHT effects $Y_{SM}$
is replaced by 
\be 
Y_s=Y_{SM} + \bar Y^{\rm even} +\frac{\bar Y_s^{\rm odd}}{\lambda_t^{(s)}}, 
\ee 
where $\bar Y^{\rm even}$ and
$\bar Y_s^{\rm odd}$ represent the effects from T-even and T-odd
particles, respectively. The branching ratio normalized to the SM
prediction is then given by 
\be
  \frac{Br(B_s\to\mu^+\mu^-)}{Br(B_s\to\mu^+\mu^-)_\text{SM}}
  = \left|\frac{Y_s}{Y_\text{SM}}\right|^2
\ee
with $Br(B_s\to\mu^+\mu^-)_\text{SM} =3.66\times 10^{-9}$.

\item[(3)] The branching ratio of $B \to X_s \mu^+\mu^-$ in the SM depends
on the functions $Y_{SM}$, $Z_{SM}$ and $D_0'(x_t)$ ($Y_{SM}$ and $D_0'$
are same as in  $B_s \to  \mu^+\mu^-$ and $B \to X_s \gamma$) and
the LHT effects enter through the modification of these functions.
The modifications of $Y_{SM}$ and $D_0'$ have been given above,
and the modification of $Z_{SM}$ is given by \cite{blht3}
\be
Z_s= Z_{SM} + \bar Z^{\rm even} + \frac{\bar Z_s^{odd}}{\lambda_t^{(s)}},
\ee
where $\bar Z^{\rm even}$ and $\bar Z_s^{odd}$ represent the effects
from T-even and T-odd particles, respectively.

\item[(4)] For the experimental values of these three B-decays, they are
given by \cite{expe}
\bea
 && Br(B \to X_s \gamma)=(3.52\pm0.23\pm0.09)\times 10^{-4}\ \
 (E_{\gamma}>1.6 {\rm GeV}), \nonumber \\
 &&Br(B_s\to\mu^+\mu^-)<7.5\times 10^{-8}, \nonumber \\
 &&Br(B \to X_s \mu^+\mu^-)=4.3^{+1.3}_{-1.2}\times 10^{-6} .
\eea
\end{itemize}
Note that throughout this work we perform our calculations in
the 't Hooft-Feynman gauge and the
SM input parameters involved are taken from \cite{pdg}.

\section{Z-boson FCNC decay $Z\to b\bar s$}
\begin{figure}[htb]
 \epsfig{file=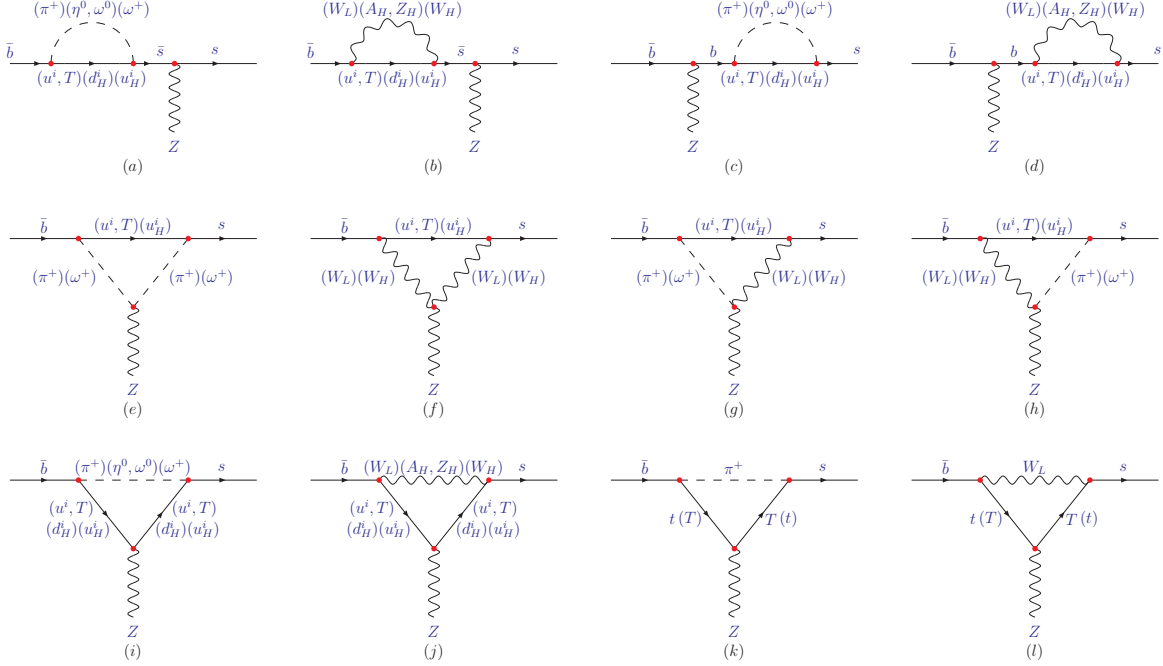,width=16cm}
 \caption{\small Feynman diagrams for $Z\to s\bar{b}$ in the LHT model.}
\label{fig1}
 \end{figure}

The relevant Feynman diagrams are shown in Fig. \ref{fig1}.
The LHT contributions are from both T-even and T-odd particles.
The contributions of T-even particles include
both the SM contributions and the contributions of the top quark T-even
partner (T-quark). The diagrams of T-odd particles are
induced by the interactions between the SM quarks and the mirror quarks
mediated by the heavy T-odd gauge bosons or Goldstone bosons.
The calculations of the loop diagrams in Fig. \ref{fig1} are
straightforward. Each loop diagram is composed of some scalar loop
functions \cite{scalarloop}, which are calculated by using LOOPTOOLS
\cite{looptool}. The relevant Feynman rules can be found in
\cite{blht3}. We have checked that the divergences of T-even
contributions are canceled at $\ord(v^2/f^2)$. For the contributions
of T-odd particles, the divergences are not canceled at
$\ord(v^2/f^2)$, which are from the diagrams with T-odd Goldstone
bosons. Such left-over divergence in the LHT model was first found
in the calculation of $B_s \to \mu^+\mu^-$ and $B \to X_s
\mu^+\mu^-$ in \cite{blht3}, and was understood as the sensitivity
of the decay amplitudes to the ultraviolet completion of the theory.
In our numerical calculations we follow \cite{blht3} to remove the
divergent term $1/\varepsilon$ and take the renormalization scale
$\mu=\Lambda$ with $\Lambda=4\pi f$ being the cutoff scale of the
LHT model.

About the involved parameters $f$ and $r$, some constraints come
from the electroweak precision data \cite{f500}, which, however,
depend on the masses of T-odd fermions and the parameter $\delta_c$
(its value is related to the details of the ultraviolet completion
of the theory). Hence, in our numerical calculations we relax the
constraints on the parameters $f$ and $r$, and let them vary in the range 
\be 
500 {\rm ~GeV}\leq f\leq 1500 {\rm ~GeV}, ~~0.5 \leq r \leq 2.0. \label{eq13}
\ee 
In addition to the parameters $f$ and $r$, the matrix
$V_{Hd}$ and the masses of $d^i_{H}$ ($i=1,2,3$) are also involved
in our calculations. To simplify our calculations, we follow
\cite{blht3} to
 consider three scenarios for these parameters:
\begin{itemize}
\item[(I)] We assume $V_{Hd} =1$ or assume 
                  the degeneracy for the masses of $d^i_{H}$, i.e.,
                  $m_{d^1_{H}}=m_{d^2_{H}}=m_{d^3_{H}}$.
                  In the former case,  we have no flavor mixing between
                  mirror down-type quarks and the SM down-type quarks
                  and thus the loop contributions of T-odd particles vanish.
                  In the latter case, due to the relation
                  of Eq.(\ref{eq5}), the masses of $u^i_{H}$ are also degenerate.
                  Then, due to the unitarity of the flavor mixing
                  matrices between mirror quarks and the SM quarks, the loop
                  contributions of T-odd particles vanish.
                  The remained contributions from the loops of T-even particles
                  depend on two parameters, i.e., the breaking scale $f$ and the
                  ratio $r$.

\item[(II)] We assume $V_{Hd} =V_{CKM}$.
In this scenario, in addition to the contributions of T-even
particles, the T-odd particles will also come into play. The
parameters involved are then $f$, $r$, $m_{d^i_H}$ and $m_{\nu^i_H}$
(the loop contributions to $B_s \to \mu^+\mu^-$ and $B \to X_s
\mu^+\mu^-$ involve the mirror lepton masses $m_{\nu^i_H}$). As
shown in Eq.(\ref{eq5}), the masses of mirror fermions are proportional to
$f$, which are assumed as
\begin{eqnarray}
&& m_{\nu^1_H}=m_{\nu^2_H}=m_{\nu^3_H}=\frac{500{\rm ~GeV}}{\rm TeV} f, \nonumber\\
&& m_{d^1_H}=m_{d^2_H}=\frac{600 {\rm ~GeV}}{\rm TeV} f, ~
m_{d^3_H}=\frac{1400 {\rm ~GeV}}{\rm TeV} f.
\end{eqnarray}
We checked that the parameters taken here satisfy the constraints from
the four-fermion interaction operators \cite{fourfermion}.

\item[(III)] We keep $\delta^d_{13}$ as a free parameter, while for other
                    parameters in the matrix $V_{Hd}$ we assume
\be 
 \delta^d_{12}=\delta^d_{23}=0,
 ~ \frac{1}{\sqrt{2}}\leq s^d_{12}\leq0.99, 
 ~ 5\times 10^{-5}\leq s^d_{23}\leq 2\times10^{-4},
 ~ 4\times 10^{-2}\leq s^d_{13}\leq 0.6. 
\ee 
For the masses $m_{d^i_H}$ and $m_{\nu^i_H}$, we take the same assumption as in
scenario II.
\end{itemize}

\begin{figure}[htb]
\epsfig{file=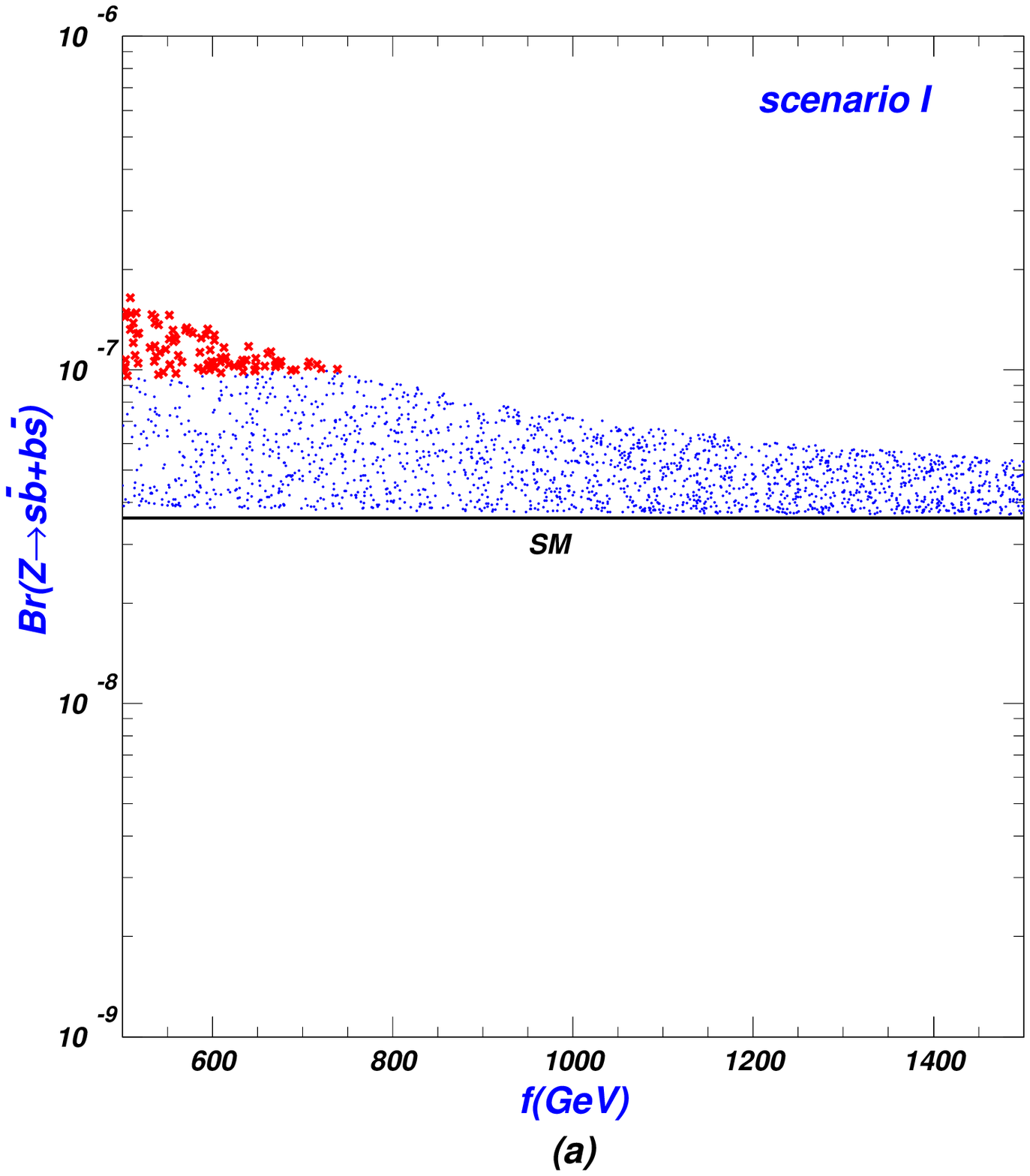,width=5.5cm}\hspace*{-0.2cm}  
\epsfig{file=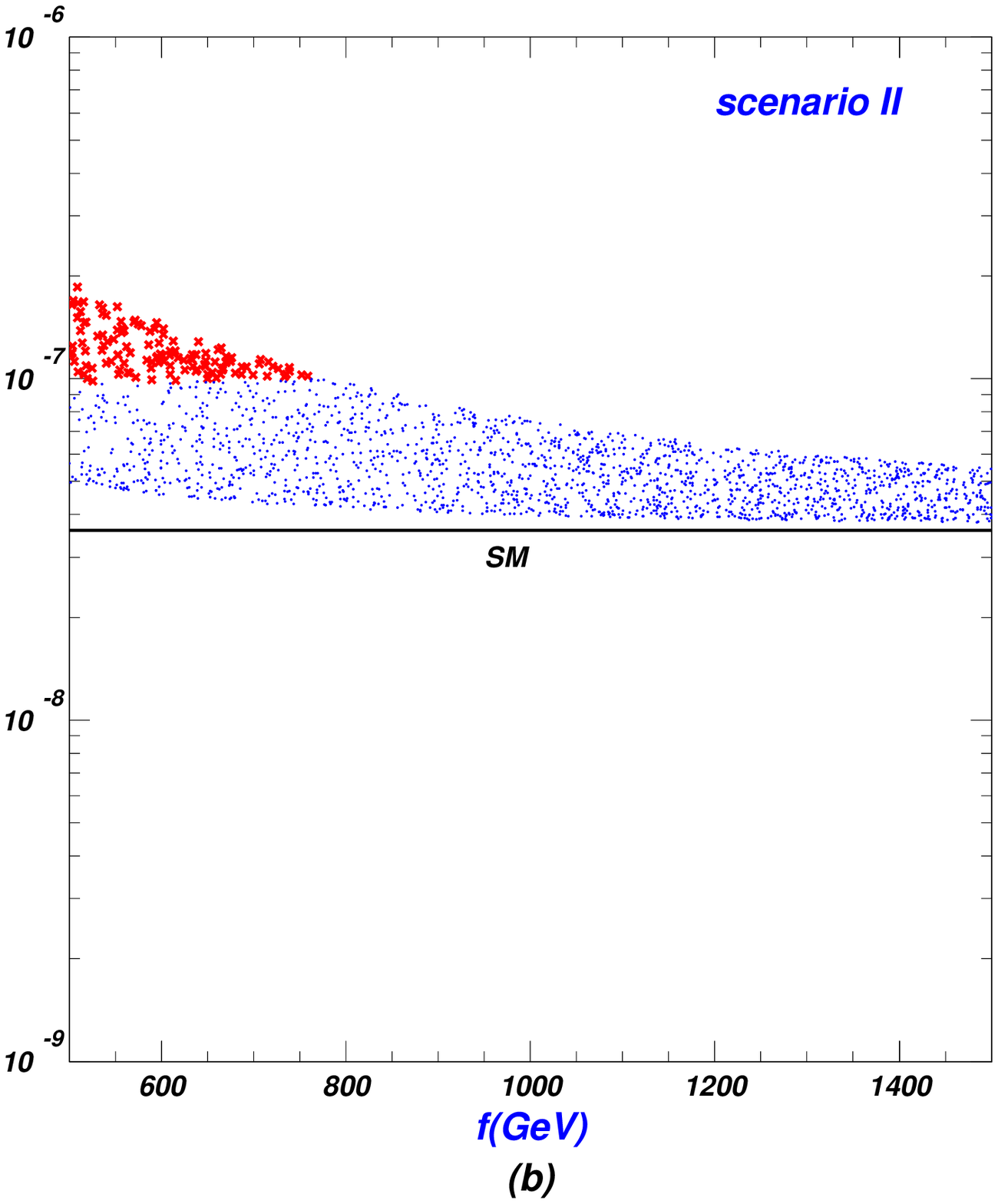,width=5.5cm}\hspace*{-0.2cm} 
\epsfig{file=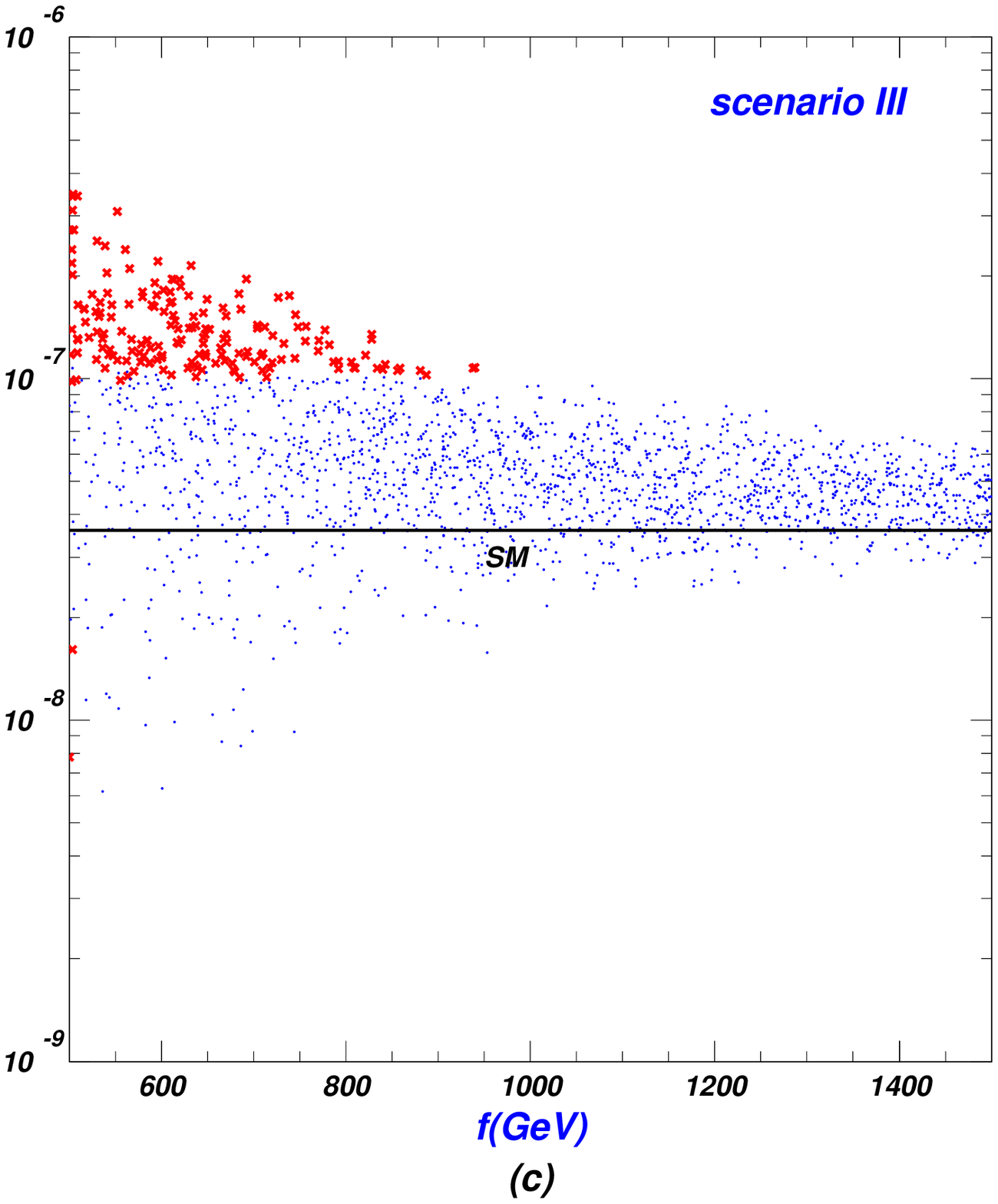,width=5.5cm} 
\vspace*{-0.5cm}  
\caption{Scatter plots for the
  branching ratio of $Z \to s\bar{b}+b\bar{s}$ versus $f$. The bullets
  (blue) and the crosses (red) are allowed and
  excluded by the $2\sigma$ B-decay constraints, respectively.}
\label{fig2}
\end{figure}
We scan over the parameters in the ranges specified above. For the
three scenarios we obtain the scatter plots in Fig. \ref{fig2},
where we also show the constraints from the B-decays $B \to X_s
\gamma$, $B_s \to \mu^+\mu^-$ and $B \to X_s \mu^+\mu^-$.

Fig. \ref{fig2} shows that the contributions are sensitive to the
scale $f$ and for lower values of $f$ the derivation from the SM
prediction is more sizable. The constraints from B-decays are
significant, with scenario-III being most stringently restrained. In
the parameter space allowed by the $2\sigma$ experimental bounds of
these B-decays, the branching ratio of $Z \to s \bar{b}+b \bar{s}$
can reach the order of $10^{-7}$,  which is about one order above
the SM prediction.

\section{Higgs boson FCNC decay $h\to b\bar s$}
The relevant Feynman diagrams involving T-even particles in the
loops can be obtained from the corresponding diagrams in Fig.
\ref{fig1} by replacing $Z$-boson with the Higgs boson. For the
contributions of T-odd particles, the diagrams are more complicate.
Note that the divergence of T-odd contributions for  $h\to b\bar s$
is  at $\ord(1)$ and is more severe than in B-decays or Z-decay
where the divergence appears at  $\ord(v^2/f^2)$. Such  divergence
is mainly due to the absence of Fig. \ref{fig1}(i) with the
down-type mirror quarks in the loops since the Higgs boson does not
couple with the down-type mirror quarks. Since such left-over
divergences in the T-odd contributions appear at $\ord(1)$, the
prediction is subject to severe theoretical uncertainty although we
can treat the divergences in the same way as for Z-decay and
B-decays discussed above. Unlike the uncertainty in Z-decay which is
correlated with the uncertainty in B-decays (and thus can be
restrained by B-decays), the uncertainty in $h \to b \bar{s}$ caused
by such T-odd contributions cannot be  constrained by B-decays since
the contributions of the diagrams mediated by the Higgs boson can be
neglected in the B-decays. To avoid such large unconstrained
uncertainty caused by T-odd contributions, we perform numerical
calculations only for scenario-I where the T-odd contributions
vanish.

\begin{figure}[htb]
\epsfig{file=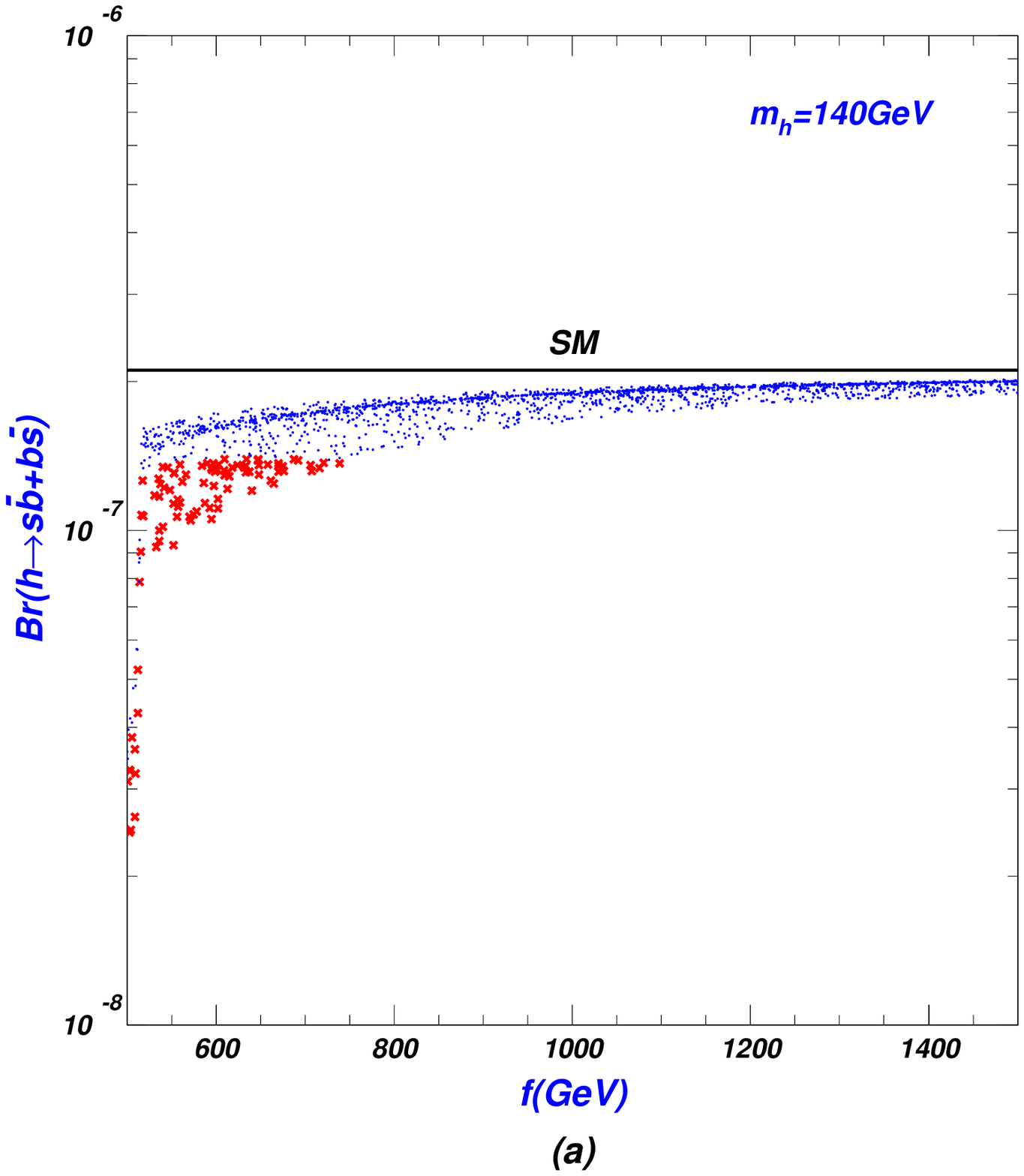,width=7cm} \hspace*{1cm}
\epsfig{file=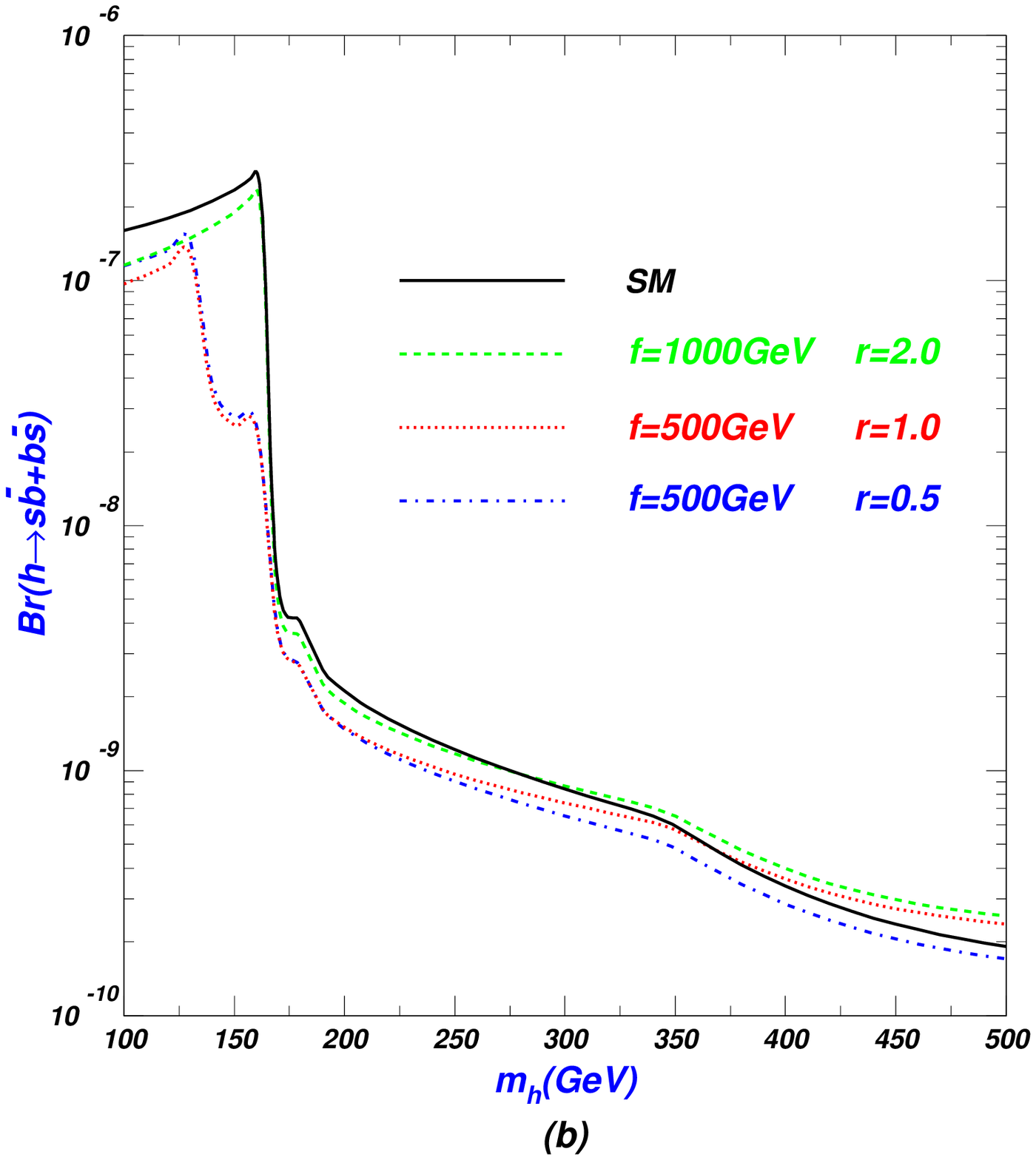,width=7cm} 
\vspace*{-0.5cm} 
\caption{\small (a) The scatter plots of $Br(h \to s\bar{b}+b\bar{s})$ versus 
          $f$ for scenario-I. The bullets (blue) and the crosses (red)
          are allowed and excluded by the $2\sigma$ B-decay constraints, respectively.
          (b) $Br(h \to s\bar{b}+b\bar{s})$ versus $m_h$ for fixed values
          of $f$ and $r$.}
\label{fig3}
\end{figure}

To evaluate the branching ratio of $h \to b\bar{s}$ we need to know
the total decay width of the Higgs boson. In addition to the decay
channels in the SM, there arise a new important channel $h\to A_{H}
A_{H}$ ($A_H$ is a candidate for the cosmic dark matter ), which may
be dominant in some parameter space of the LHT model \cite{htoah}.
The total decay width is given by 
\be
  \Gamma_{total}\approx\Gamma_{h\to {\rm fermions}}
  +\Gamma_{h\to W_{L}W_L}+\Gamma_{h\to Z{_L} Z{_L}}+
   \Gamma_{h\to A{_H} A{_H}} .
\ee
In Fig. \ref{fig3} (a) we scan over $r$ and $f$ in the ranges in
Eq.(\ref{eq13}) and present the scatter plots for the branching ratio with
$m_h =140$ GeV. In Fig. \ref{fig3} (b) we show the dependence of the
branching ratio on the Higgs boson mass by fixing the parameters $r$
and $f$ allowed by the electroweak precision data and B-decays. In
our calculations we kept the order up to $\ord(v/f)$ and checked
that the divergences are canceled to this order.

From Fig. \ref{fig3} we see that the branching ratio in the LHT
model is below the SM prediction and the deviation is significant
for low values of $f$. In the parameter space allowed by the
B-decays at $2\sigma$ level, the branching ratio can be one order
below the SM prediction due to the fact that in some parameter space
the decay $h\to A{_H} A{_H}$ may be dominant and greatly enhance the
total width.

\section{Conclusion}
The littlest Higgs model with T-parity may have flavor problem since
it predicts new flavor-changing interactions
between mirror fermions and the Standard Model fermions, which can
induce various FCNC decays.
Since all these decays induced in this model are correlated,
we in this work performed a collective study
 for the FCNC decays of B-mesons, Z-boson and Higgs boson.
We found that under the current experimental constraints from the B-decays,
the branching ratios of both $Z \to b \bar{s}$ and $h \to b \bar{s}$
can still deviate from the SM predictions significantly.
In the parameter space allowed by the B-decays, the branching ratio of
$Z \to b \bar{s}$ can be enhanced to $10^{-7}$ (about one order above the
SM prediction) while $h\to b \bar{s}$ can be much suppressed relative to
the SM prediction (about one order below the SM prediction).

We remark that unlike the FCNC B-decays, it is quite challenging to
test these rare Z-boson and Higgs boson decays at collider
experiments. For instance, to test this rare decay of Z-boson, we
may need the GigaZ option of the ILC. Theoretically, for the test of
the LHT model, these rare decays are complementary to the direct
production of T-quark \cite{T-quark-LHC} and the production of top
quark or Higgs boson  \cite{wang} whose cross sections can be
sizably altered by the LHT model.

\section*{Acknowledgement}
This work was supported in part by National Natural
Science Foundation of China (NNSFC) under grant No. 10725526 and 10635030.

\end{document}